\documentclass[jap,aip,english,onecolumn,notitlepage,superscriptaddress]{revtex4-2}
\usepackage[T1]{fontenc}  
\usepackage[latin1]{inputenc}
\usepackage{multirow,amsmath}
\usepackage{babel} 
\usepackage{txfonts}
\usepackage{epsfig,epsf,psfrag} 
\usepackage{graphicx} 
\usepackage{subfigure}
\usepackage{pslatex}
\usepackage{float}
\usepackage{upgreek}
\usepackage{subfigure}
\usepackage{epic,eepic} 
\usepackage{color,pstcol}
\usepackage{pstricks} 
\usepackage{fancyhdr} 
\usepackage{tabularx}
\usepackage{physics}
\usepackage{url}
\usepackage{listings}
\usepackage{color}
\newcommand{\RN}[1]
{%
  \textup{\uppercase\expandafter{\romannumeral#1}}%
}

\begin{document}
\title{A thermodynamic probe of the topological phase transition in epitaxial graphene based Floquet topological insulator}
\author{Abhishek Kumar}\email{abhi.kumar@niser.ac.in}\altaffiliation{Current address: Department of Physics, University of Massachusettes, Amherst, MA 01003, USA}
\affiliation{School of Physical Sciences, National Institute of Science Education \& Research, HBNI, Jatni-752050, India} 
\author{Colin Benjamin}\email{colin.nano@gmail.com (Corresponding Author)}
\affiliation{School of Physical Sciences, National Institute of Science Education \& Research, HBNI, Jatni-752050, India}  
\begin{abstract}
One can use light to tune certain materials, from a trivial to a topological phase. A prime example of such materials, classified as Floquet topological insulators (FTI), is epitaxial graphene. In this paper, we probe the topological phase transition of {a} FTI via the efficiency and work output of quantum Otto and quantum Stirling heat engines. A maximum/minimum in the efficiency or work output invariably signals the phase transition point. Further, both engines' work output and efficiency are markedly robust against the polarization direction of light.
\end{abstract}
\maketitle
\section{Introduction}
    Cyclic quantum heat engines consists of quantum working substance, hot and cold thermal baths, and different quantum thermodynamic strokes  like quantum isochoric wherein energy levels of the quantum system remain unchanged, quantum adiabatic in which the occupation probability of the distinct energy levels remain unchanged and quantum isothermal, which is similar to its classical counterpart with the temperature remaining constant, see Refs.~\cite{quan2007quantum,quan2009quantum}. Not only quantum heat engines show an increase in power output or can be tuned to be more efficient than classical heat engines~\cite{PhysRevE.97.042120,klatzow2019experimental}, they can also have novel applications in probing phase transitions \cite{ma2017quantum}. {Furthermore, several interesting proposals~\cite{wertnik2018optimizing,zagoskin2012squeezing,funo2019speeding} and experimental realizations~\cite{ono2020analog,guthrie2021cooper} are present in literature for cyclic quantum heat engines.}~Motivated by this in a recent work, trivial to topological phase transition was probed via a {cyclic} quantum heat engine cycles\cite{fadaie2018topological}. Since Floquet engineering and the dressing of band structure using external lasers has attracted great deal of interest~\cite{de2019floquet,bukov2015universal}. We in this work look at the efficiency and work output of quantum heat engines as a probe for the phase transition from trivial phase to a topological phase in a FTI.~Specifically, we discuss the ability of the quantum heat engine to probe the topological phase transition point at both low and high temperature regimes of an epitaxial graphene layer, our modeled FTI. However, there are some lacuna in using quantum heat engine cycles as has also been seen in Ref.~\cite{fadaie2018topological}.~The quantum Stirling cycle wasn't able to probe the topological phase transition in the {high} temperature regime. Further the work output and efficiency at {high} temperatures lose the symmetry about the phase transition point implying that different topological phases may influence the work output and efficiency for the two engines cycle.~Motivated by this we not only provide a template for probing the topological phase transition in a FTI via a quantum heat engine cycle, we also ameliorate the deficiency of the previous attempt\cite{fadaie2018topological}.
   
   The effect of off-resonant light on both silicene family (silicene, germanene, and stanene)~\cite{ezawa2013photoinduced} and epitaxial graphene~\cite{cayssol2013floquet,PhysRevB.84.235108,zhai2014photoinduced} has been studied using Floquet theory in the high-frequency regime~\cite{bukov2015universal}.~Off-resonant light modifies the band and lifts valley degeneracy, both for silicene family and for epitaxial graphene.~{However external static electric field preserves valley degenracy in silicene family.~Hence, band structure for stanene in presence of static electric field differs from epitaxial graphene in presence of off-resonant light. This difference in band structure is explored to construct the quantum Otto engine(QOE) and quantum Stirling engine(QSE) which form a better probe for the topological phase transition point as compared to the case of stanene in presence of static electric field~\cite{fadaie2018topological}}. Further, this work is also the first time anyone has probed the topological phase transition point of a FTI. Our aim here is threefold, 1) To probe the topological phase transition and look at how the different topological phases affect the work done and efficiency of QOE and QSE in a FTI, 2) To check how good our heat engine is as a topological phase transition probe in different temperature regimes, 3) Finally, we would like to check whether the topological quantum heat engine with the FTI as working substance is robust to polarization of light?   
  
   The summary of the findings of this paper are as follows, 1) we see that the work output and efficiency are independent of the phase (in topological or not) of the FTI, 2) The topological phase transition point can be effectively probed in both low and high temperature regimes, and 3) The work output and efficiency calculated from both QOE and QSE with FTI as the working substance are robust to the polarisation of light.
   
   The outline of this paper is as follows, we first discuss the theory of topological phases in a Floquet topological insulator(FTI) based on epitaxial graphene. Next, we discuss the quantum heat engines proposed to probe the topological phase transition point in the FTI. In section $\RN{3}$, we present the results obtained for both QOE and QSE with FTI as the working substance. We conclude in the last section with a summary of the results.
   
\section{Theory}
  \subsection{Floquet Hamiltonian: Band structure and Chern number}
   The monolayer silicene family share buckled shape structure and have a spin-orbit term which appears as the Dirac mass term in their low energy effective Hamiltonian.~The mass term opens a gap at the two Dirac points and hence these behave as an insulator with the valley degeneracy remaining intact~\cite{ezawa2015monolayer}. One can close the gap at both the Dirac points by applying an electric field. Though graphene is made up of an element belonging to the same group in periodic table as that of other silicene family members, the structure of graphene is planar and hence there is no mass term due to the absence of spin-orbit coupling which results in closing of gap at the two Dirac points and hence we get a semimetal.~However, in epitaxial graphene, the substrate potential creates a site energy difference between two sublattices which leads to an additional mass term that opens a gap at the Dirac points but the valley degeneracy is preserved.
   
\begin{figure}[H]
	\begin{center}
	    \includegraphics[width=8.25cm, height = 5.5cm]{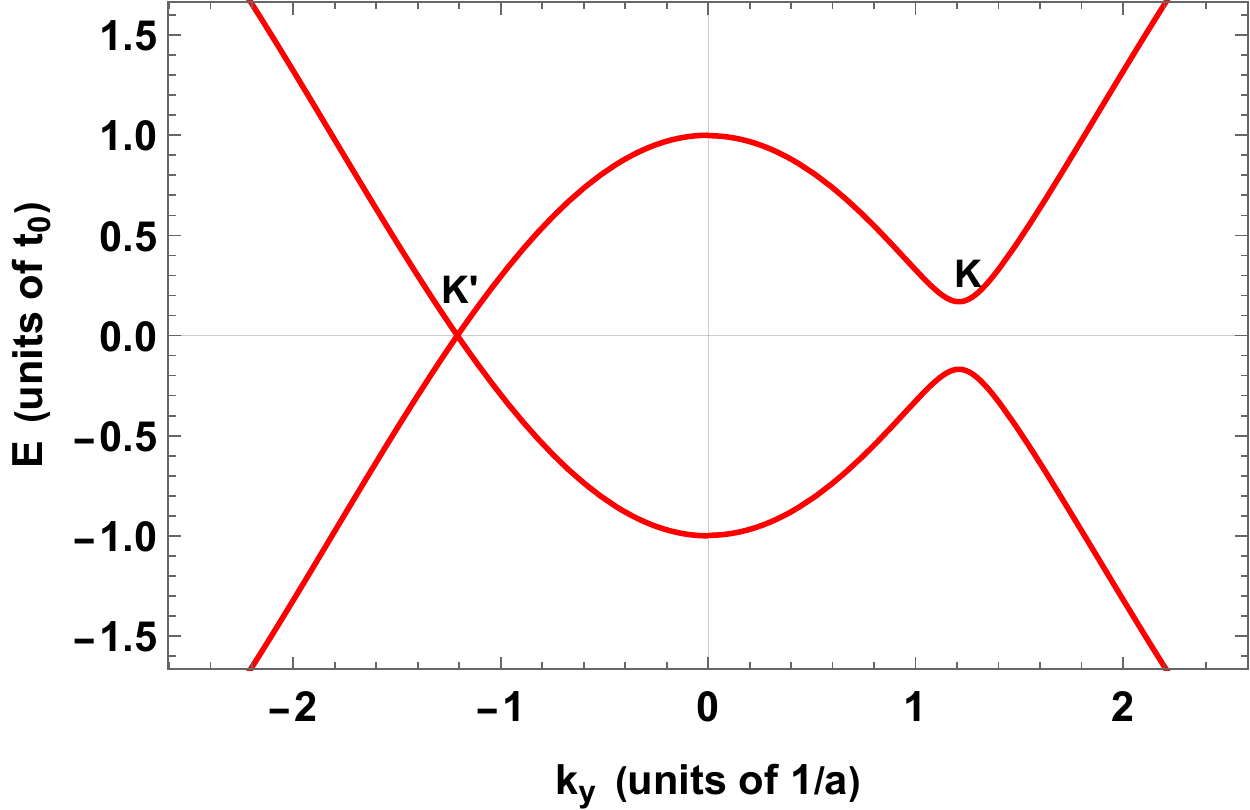}
	\end{center}
	\caption{Band structure plot (red) with energy($E$) given in units of $t_{0}$ at $\Delta = I_{+} = 0.1t_{0}$ and $a$ is the nearest neighbor distance($E$ vs $k_{y}$). The lattice constant is equal to $\sqrt{3}a = 2.46$ \mbox{\AA }~\cite{ezawa2015monolayer} for epitaxial graphene. The polarization of light will affect the spectrum, we drew the spectrum for right polarized light but if one uses left circularly polarized light then the gap closes for $K$ point and remains open at $K^{\prime}$. The gap at $K$ and $K^{\prime}$ is not same which means that there is no valley degeneracy.}
	\label{fig:spectrum}
    \end{figure}

\begin{figure}[H]
	\begin{center}
	    \includegraphics[width=8cm, height = 6cm]{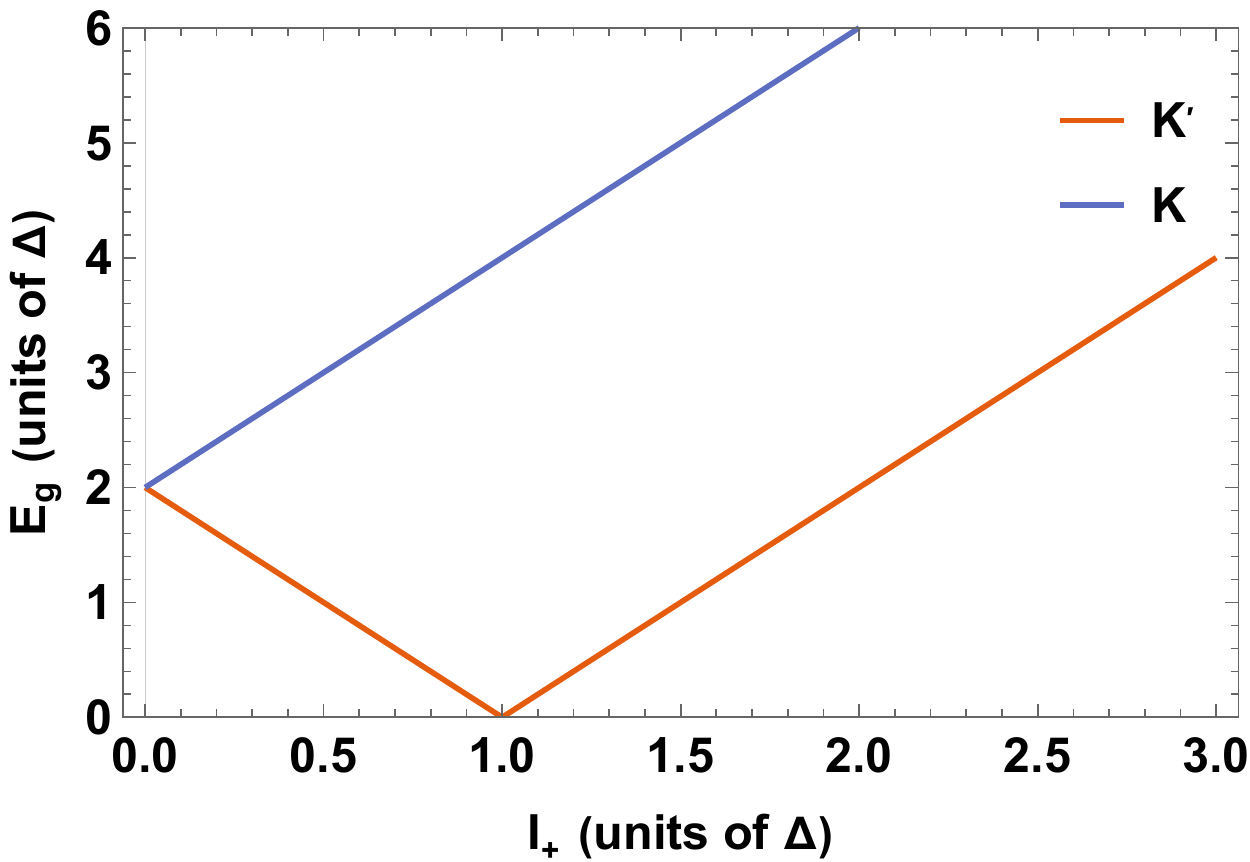}
	\end{center}
	\caption{ Band gap plotted at $\Delta = 0.1t_{0}$ for right circularly polarized light ($E_{g}$ vs $I_{+}$). The gap closes at $K^{\prime}$ (orange)  when $I_{+} = \Delta$ but remains open at $K$ (blue). Both gap functions vary linearly with $I_{+}$ and are separated at all values of $I_{+}$ except the point where photoillumination parameter vanishes. This clearly shows that gap at two Dirac points are inequivalent.}
	\label{fig:energygap}
    \end{figure}

   We first discuss the Floquet topological insulator seen in an epitaxial graphene substrate~\cite{novoselov2004electric} irradiated with off-resonant circularly polarized light~\cite{zhai2014photoinduced}.~The low energy effective Hamiltonian for a 2D epitaxial graphene substrate lying in the $xy$ plane is given as
   \begin{eqnarray}
       H_{\eta}(k_{x},k_{y}) = \hbar v_{F}(\sigma_{x}k_{x} + \eta\sigma_{y}k_{y}) + \Delta\sigma_{z},
       \label{eqn:Graphene_Hamiltonian}
   \end{eqnarray}
  where $\eta$ is $\pm 1$ for the Dirac $K$ and $K^{\prime}$ points.~The momentum $k_{x}$, $k_{y}$ are taken about the $K$ and $K^{\prime}$ points, $v_{F}( = 3at_{0}/\hbar)$ is Fermi velocity where the parameters $a$ and $t_{0}$ in parenthesis denote nearest neighbor distance and hopping amplitude in graphene. $\Delta$ denotes substrate potential~\cite{Zhou_2007,Zhou_2008} which creates $2\Delta$ site energy difference between the two sublattices.~The Hamiltonian in Eq.~\ref{eqn:Graphene_Hamiltonian} is same for both spins ($\uparrow,\downarrow$). When irradiated with off-resonant circularly polarized light with vector potential $\textbf{A} = (A\chi\sin (\omega t), A\cos (\omega t))$ where $\chi = \pm 1$ is for right(left) circularly polarized light.~The spatial dependence is dropped in vector potential term since wavelength of light used is much larger as compared to sample size.~This light field shifts momentum vector $\textbf{k} \rightarrow \textbf{k} + \frac{e\textbf{A}}{\hbar}$ in Eq.~\ref{eqn:Graphene_Hamiltonian}, where $e$ is the electron charge.~The Floquet theory can be applied to new Hamiltonian since each component of the vector potential is time periodic. Using high frequency expansion in Floquet picture~\cite{bukov2015universal}, the perturbative first order Floquet Hamiltonian is given as
   \begin{eqnarray}
      H_{F,\eta}(\textbf{k}) &=& H_{0,\eta}(\textbf{k}) + \sum_{l=1}^{\infty}\frac{[H_{-l,\eta},H_{l,\eta}]}{\hbar\omega},
      \label{eqn:Floquet_Hamiltonian_Appendix}
  \end{eqnarray}
  where $H_{l,\eta}$ for all integer $l$ values are defined as
  \begin{eqnarray}
      H_{l,\eta}(\textbf{k}) = \frac{1}{T}\int^{T}_{0}\exp (il\omega t)H_{\eta}(\textbf{k},t) dt.
      \label{eqn:fourier_transform_Appendix}
  \end{eqnarray}
    with $T(=2\pi/\omega$) being the time period of the Hamiltonian and $l$ takes integer values, and $H_{\eta}(k,t)$ inside the integral is the Hamiltonian obtained after replacing the momentum vector ($\textbf{k}$) with the new momentum ($\textbf{k} + e\textbf{A}/\hbar$).~The first term of Eq.~\ref{eqn:Floquet_Hamiltonian_Appendix} is obtained using Eq.~\ref{eqn:fourier_transform_Appendix} by taking $l = 0$, i.e.,
   \begin{eqnarray}
       H_{0,\eta}(\textbf{k}) = \frac{1}{T}\int_{0}^{T}H_{\eta}(\textbf{k},t)dt,
       \label{eqn:first_order_term}
   \end{eqnarray}
   with 
   \begin{eqnarray}
       H_{\eta}(\textbf{k},t) &=& \hbar v_{F}\left[\sigma_{x}\left(k_{x} + eA_{x}/\hbar\right) + \eta\sigma_{y}\left(k_{y} + eA_{y}/\hbar\right) \right]\nonumber \\&+& \Delta\sigma_{z},
       \label{eqn:Floquet_time_dependent_Hamiltonian}
   \end{eqnarray}
   where both $A_{x} = \chi A\sin (\omega t)$ and $A_{y} = A\cos (\omega t)$ are periodic with time $T$, when we perform the integral in Eq.~\ref{eqn:first_order_term} we obtain no contribution to the integral from terms containing component $A_{x}$ and $A_{y}$ and the other terms with no time dependence remain the same after the integration. Thus, we get
   \begin{eqnarray}
       H_{0,\eta}(\textbf{k}) = H_{\eta}(k_{x},k_{y}),
       \label{eqn:first_order_result}          
   \end{eqnarray}
   which is the Hamiltonian obtained in Eq.~\ref{eqn:Graphene_Hamiltonian} with no light field. The sum in second term of Eq.~\ref{eqn:Floquet_Hamiltonian_Appendix} is limited to $l=1$ since the terms with $l \geq 2$ vanish as both $\sin (\omega t) = (\exp (i\omega t) - \exp (-i\omega t))/2i$ and $\cos (\omega t) = (\exp (i\omega t) + \exp (-i\omega t))/2$ has contribution from $l =1$ and $-1$, thus all higher Harmonics with $l \geq 2$ in Eq.~\ref{eqn:fourier_transform_Appendix} vanishes. This gives 
   \begin{eqnarray}
       \sum_{l=1}^{\infty} [H_{-l,\eta},H_{l,\eta}] = [H_{-1,\eta},H_{1,\eta}],
       \label{eqn:sum}
   \end{eqnarray}
    where $H_{-1,\eta}$ and $H_{1,\eta}$ are obtained by using Eq.~\ref{eqn:fourier_transform_Appendix} and Eq.~\ref{eqn:Floquet_time_dependent_Hamiltonian}. Solving for them, we get
    \begin{equation}
        H_{\pm1,\eta} = \frac{eAv_{F}}{2}\left(\pm\chi i\sigma_{x} + \eta\sigma_{y}\right).
        \label{eqn:H_{1}}
    \end{equation}
    Calculating the commutator given in (\ref{eqn:sum}), we get  $[H_{-1,\eta},H_{1,\eta}]/\hbar\omega = \eta I_{\chi}(A)\sigma_{z}$ where $I_{\chi}=\chi(eAv_{F})^{2}/\hbar\omega$. This gives 
    \begin{eqnarray}
      H_{F,\eta}(\textbf{k}) = H_{\eta}(k_{x},k_{y}) + \eta I_{\chi}(A)\sigma_{z}.
      \label{eqn:quasisteady_Hamiltonian}
  \end{eqnarray}
  In matrix representation, we have
    \begin{equation}H_{F,\eta}(\textbf{k}) = 
    \begin{pmatrix}
    \Delta + \eta I_{\chi}&  \hbar v_{F} (k_{x} - i\eta k_{y}) \\
    \hbar v_{F}(k_{x} + i\eta k_{y}) &-(\Delta + \eta I_{\chi}) \\
  \end{pmatrix},\end{equation}
   where $I_{\chi} = \chi(eAv_{F})^{2}/\hbar\omega$ is a measure of Intensity and is defined as photo illumination parameter.~The second term in Eq.~\ref{eqn:quasisteady_Hamiltonian} is the change due to light field which adds up to previous mass term and it depends on the amplitude of vector potential.~When off-resonant light is used it modifies electron band structure by virtual photon absorption processes and the off-resonant condition is satisfied at frequency $\omega \gg t_{0}/\hbar$, where $t_{0}$ is the nearest neighbor hopping amplitude in graphene and it is equal to $3 eV$. Thus, the lowest frequency at which off-resonant condition is satisfied is $10^{15}$ Hz. If we write the Hamiltonian in Eq.~\ref{eqn:quasisteady_Hamiltonian} in the form $\vec{\textbf{d}}\cdot\vec{\sigma}$, we get $d_{x} = \hbar v_{F} k_{x}$, $d_{y} = \eta\hbar v_{F}k_{y}$, and $d_{z} = \Delta + \eta I_{\chi}$ and the eigenvalues are 
  \begin{eqnarray}
      E_{\eta}^{\sigma}(\textbf{k}) = \sigma|\vec{d} | = \sigma\sqrt{(\hbar v_{F} k)^{2} + (\eta I_{\chi} + \Delta)^{2}}
      \label{eqn: eigenvalues}
  \end{eqnarray}
  where $\sigma = \pm$, $k = |\textbf{k}|$ and $\pm$ sign is for conduction and valence band respectively. The spectrum is plotted in Fig.~\ref{fig:spectrum} for $\Delta = I_{+} = 0.1t_{0}$ where subscript $+$ in $I$ stands for right circularly polarized light. It is clear from Fig.~\ref{fig:spectrum} that the valley degeneracy is lifted. 
    To know the gap at $K$ and $K^{\prime}$ we set $k = 0$ in Eq.~\ref{eqn: eigenvalues} and subtract the eigenvalue of valence band from conduction band. This gives
    \begin{eqnarray}
        E_{g,\eta} = 2|\eta I_{\chi} + \Delta|.
        \label{eqn:energygap}
    \end{eqnarray}
    The polarization of light (right or left) affects the spectrum as it interchanges $K$ and $K^{\prime}$ in Fig.~\ref{fig:spectrum}, i.e., the gap closes for $K$ in case of left circularly polarized light while it closes for $K^{\prime}$ in case of right circularly polarized light. The gap function in Eq.~\ref{eqn:energygap} is shown at $\Delta = 0.1t_{0}$ in Fig.~\ref{fig:energygap}.
    
\begin{figure}
    \begin{center}
	    \includegraphics[width=8cm, height = 8cm]{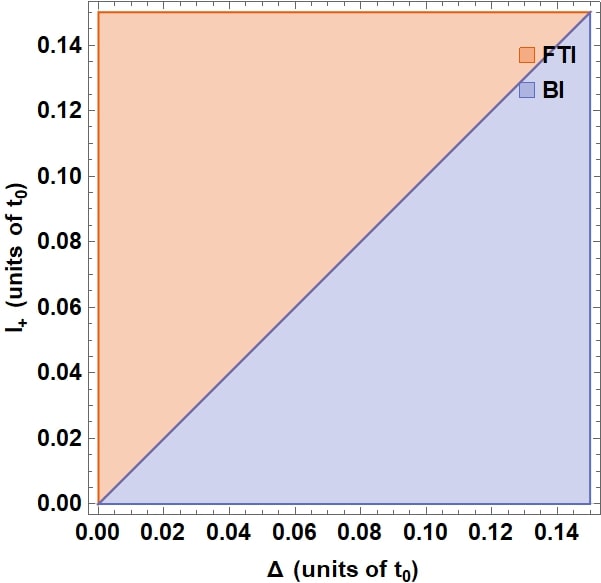}
	\end{center}
	\caption{Topological phase diagram which distinguishes the FTI(orange) and band insulator(blue) ($I_{+}$ vs $\Delta$). The boundary between them is the place where gap closes at $K^{\prime}$ point. For left circularly polarized light the phase diagram remains same except $I_{+}$ is replaced by $I_{-}$. The plot shows that a threshold value of photo illumination parameter($I$) is always present for all values of $\Delta$ where the topological phase change occurs.}
	\label{fig:topological_phase_diagram}
    \end{figure}

    One can explore different topological phases by calculating the Berry curvature and Chern number for FTI Hamiltonian based on epitaxial graphene as in Eq.~\ref{eqn:quasisteady_Hamiltonian}.~The local Berry curvature~\cite{ezawa2015monolayer} is given by
    \begin{eqnarray}
        \Omega_{\eta} (\textbf{k}) = \pm\frac{1}{2}\frac{\vec{d}(\textbf{k})}{|\vec{d}(\textbf{k})|^{3}}\cdot(\partial_{k_{x}}\vec{d} \times \partial_{k_{y}}\vec{d} )
        \label{eqn:formula_berry_curvature}
    \end{eqnarray}
    where $\pm$ sign is for conduction and valence band.
     Substituting $\vec{d}$ for the Floquet Hamiltonian as obtained before, we get
    \begin{eqnarray}
        \Omega_{\eta} (\textbf{k}) = \frac{(\hbar v_{F})^{2}(\eta\Delta + I_{\chi})}{2((\Delta + \eta I_{\chi})^{2} + (\hbar v_{F}k)^{2})^{3/2}}
        \label{berry_curvature_Floquet_Hamiltonian}
    \end{eqnarray}
    This is the  Berry curvature for the conduction band. The Berry curvature for valence band is $-\Omega_{\eta} (\textbf{k})$.
    The Chern number is defined as an integral of Berry curvature over the 2D Brillouin zone, i.e.,
    \begin{eqnarray}
        C_{\eta} = \frac{1}{2\pi}\int d^{2}k   \Omega_{\eta} (\textbf{k}).
        \label{eqn:chern_number_formula}
    \end{eqnarray}
    On integration, we get the Chern number of conduction band to be
    \begin{eqnarray}
        C_{\eta} = \frac{\eta}{2}sgn(\Delta + \eta I_{\chi}),
        \label{eqn:chern_number_eqpitaxial_graphene}
    \end{eqnarray}
    and for valence band we just have an extra minus sign since the berry curvature is replaced by $-\Omega_{\eta} (\textbf{k})$. From Eq.~\ref{eqn:chern_number_eqpitaxial_graphene} it is seen that tuning either sublattice potential($\Delta$) or polarisation $\chi$ or the photo illumination parameter $I$ will enable us to see different topological phases. We fix the polarization direction($\chi$) to be right and compute the total Chern number 
    \begin{eqnarray}
        C = C_{K} + C_{K^{\prime}}.
        \label{eqn:total_chern_number}
    \end{eqnarray}
    We observe two distinct phases for $C =1$ when $\Delta < I_{+}$ and $C = 0$ when $\Delta > I_{+}$. $C = 1$ represents FTI \cite{zhai2014photoinduced} where we get gapless edge state under open boundary condition and $C = 0$ is the trivial band insulator(BI). We did the Chern number calculation for conduction band with right circularly polarized light. If the polarization direction is chosen to be left then the conduction band Chern number becomes -1 (FTI) when $\Delta < I_{-}$ and 0 (BI) when $\Delta > I_{-}$. The topological phase diagram is plotted in Fig.~\ref{fig:topological_phase_diagram} for right polarized light.~Now, we know the existing topological phases in a photoinduced epitaxial graphene based on the Chern number calculation.~Below we devise QHEs based on epitaxial graphene which identify the transition point from trivial to topological phase.
    
  \subsection{Basics of quantum Otto and quantum Stirling engine} 
  Cyclic quantum heat engines consists of quantum working substance, hot and cold thermal baths, and different quantum thermodynamic strokes.~Based on the combination of different quantum thermodynamic strokes in each cycle we can get many different types of quantum heat engines, see Refs.~\cite{quan2007quantum,quan2009quantum}. We will focus on work output and efficiency of quantum Otto and quantum Stirling cycles which are both $4$-stroke cycles with the FTI based on epitaxial graphene as the quantum working substance.~The energy eigenvalues of the FTI with epitaxial graphene substrate are given by $E_{\eta}^{\sigma}(\textbf{k})$ see Eq.~\ref{eqn: eigenvalues}. As seen from Fig.~\ref{fig:spectrum}, there are only two bands and conduction band energy is negative of valence band energy. Further, two valleys are non degenerate. Before starting the calculation of work output and efficiency, let us review the basic theory of quantum heat engines.~{The internal energy~\cite{munoz2016magnetically,quan2007quantum,pathria32statistical} is given by $U = Tr(\rho H)  = \sum_{n}P_{n}E_{n}$, where $\rho$ is the density operator and $P_{n}$ is the occupation probability of the  $n^{th}$ eigenstate with energy eigenvalue $E_{n}$.~Further, from first law of thermodynamics, $dU = \text{\dj}Q + \text{\dj}W =  \sum_{n}E_{n}dP_{n}+P_{n}dE_{n}$, where the two terms in the sum are in correspondence with the macroscopic notions of heat and work respectively. Further, for system in thermal equilibrium we can use the definition of heat $\text{\dj}Q = TdS$~\cite{quan2007quantum}, where $T$ is the temperature of the heat bath placed in thermal equilibrium with the open quantum system and $S$ is the entropy.~The entropy can be written in terms of $P_{n}$ as $S = -k_{B}\sum_{n}(P_{n}\ln P_{n} + (1-P_{n})\ln (1-P_{n}))$. These definitions will further get generalized below for the Floquet topological insulator with sum over their respective quantum labels.} After evaluating the trace of Internal energy, we get a weighted sum of energy eigenvalues and the associated weights are nothing but the Fermi function. The normal sum over energy quantum numbers will get replaced by an integral over $\textbf{k}$ as $\textbf{k}$ varies continuously in the momentum space~\cite{fadaie2018topological}. This gives   
    \begin{equation}
        U = \sum_{\eta,\sigma}\int \frac{d^{2}\textbf{k}}{(2\pi)^{2}} g(E_{\eta}^{\sigma})f(E_{\eta}^{\sigma},T)E_{\eta}^{\sigma}(\textbf{k})
        \label{Internal_energy}
    \end{equation} 
    where $g(E_{\eta}^{\sigma})$ is degeneracy of given eigenvalue, $f(E_{\eta}^{\sigma},T)$ is occupation probability of energy eigenvalue $E_{\eta}^{\sigma}(\textbf{k})$ and is given by Fermi function
    \begin{eqnarray}
        f(E_{\eta}^{\sigma}(\textbf{k}),T) = \frac{1}{\exp ((E_{\eta}^{\sigma}(\textbf{k})-E_{F})/k_{B}T) + 1}
        \label{eqn:fermi_function}
    \end{eqnarray}
    where $k_{B}$ is the Boltzmann constant and $T$ denotes temperature. The $\eta$ and $\sigma$ summing index as per definition runs over $\{K,K^{\prime}\}$ and $\{+,-\}$ respectively. Thus, Eq.~\ref{Internal_energy} is sum of conduction and valence band internal energy, i.e., $U = U^{+} + U^{-}$ where $+$ and $-$ indicates positive and negative band. In case of epitaxial graphene, Fermi energy $E_{F}$ is zero, spin degeneracy gives $g(E_{\eta}^{\sigma}) = 2$ and band structure obeys the relation $E_{\eta}^{+}(\textbf{k}) = - E_{\eta}^{-}(\textbf{k})$ where superscript $+$ and $-$ stands for conduction and valence band respectively. As we restrict ourselves to low temperature regime where large $k$ values do not contribute to the integral, the integral limit of $k$ from $0$ to $\infty$ in Eq.~\ref{Internal_energy} works well even on using low energy band spectrum(hold true in a radius $k\sim 1/\sqrt{3}a$) instead of full energy spectrum.
    
    But we still encounter a divergence in Eq.~\ref{Internal_energy} for $\sigma = -$ term. To overcome this divergence one uses simple re-normalization where internal energy is subtracted by the zero temperature ground state energy($= \sum_{\eta}\int d^{2}\textbf{k}E_{\eta}^{-}(\textbf{k})$) as also discussed in Ref.~\cite{fadaie2018topological}. This gives $U = 2U^{+}$ and $U^{+}$ is given by
    \begin{eqnarray}
        U^{+} = \sum_{\eta}\int d^{2}\textbf{k} f(E_{\eta}^{+},T)E^{+}_{\eta}(\textbf{k}) \label{renormalized_internal_energy}
    \end{eqnarray}
    and it only depends on the positive energy spectrum. Thus, this limits our calculation of work output and heat flow for only positive bands as $dU = \text{\dj}Q + \text{\dj}W$ (first law of thermodynamics). The total work output and  heat is given as twice of the contribution coming from the positive band. This is how we take care of divergence from the integral of negative band.~Several studies which are related to the thermodynamics of graphene use only the positive bands for calculation of partition function and ignore the negative contribution, see~\cite{pena2015magnetostrain,boumali2015thermodynamic}.  
    
    In the next two subsections we will compute the work output and efficiency of QOE and QSE with epitaxial graphene irradiated by an off-resonant circularly polarized light as our quantum working substance.~The substrate potential $\Delta$ was determined experimentally to be of the order of few $100 meV$ by previous studies using epitaxial graphene~\cite{zhou2008origin,rotenberg2008origin} and can be tuned artificially. We for our convenience fix the substrate potential $\Delta = 0.1t_{0} = 300 meV$ and direction of polarization of light is chosen to be right circularly polarized($\chi = +$). We can do the same calculation for $\chi = -$ and all the results in the subsequent section holds for $\chi = -$ too. We justify in section $\RN{3}$A and $\RN{3}$B that the same result will be obtained when $\chi = -$. Therefore our results are independent of the polarization of circularly polarized light. The cycle of the engines are run by varying the photo illumination parameter $I_{+}$ and using two thermal baths kept at different temperatures.

  \subsubsection{\rm{\textbf{Quantum Otto engine cycle}}}
\begin{figure}[H]
    	\begin{center}
    		\includegraphics[width=7.25cm, height = 6.25cm]{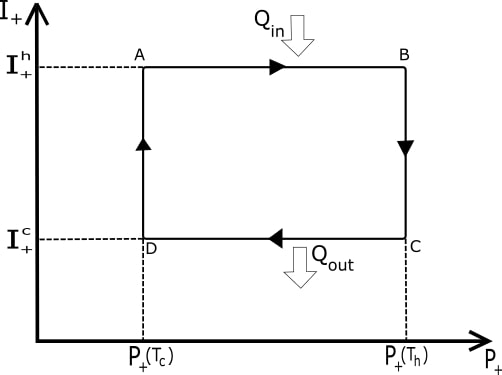}
    	\end{center}
    	\caption{ The cycle of QOE has two isochoric and adiabatic processes from $A\rightarrow B$, $C \rightarrow D$ and from $B\rightarrow C$, $C \rightarrow D$ respectively. The heat energy input in the $A \rightarrow B$ stage is $Q_{in}$ and $Q_{out}$ is the heat energy output in the $C\rightarrow D$ stage. The hot and cold bath temperatures for the QOE cycle are set at $T_{h}$ and $T_{c}$ respectively. The illumination parameter is varied between $I_{+}^{c}$ and $I_{+}^{h}$ with corresponding energy $E_{\eta}^{+,c}$ and $E_{\eta}^{+,h}$. The occupation probability take values between $P_{+}(T_{c}) = f(E_{\eta}^{+,c}(\textbf{k}),T_{c})$ and $P_{+}(T_{h}) = f(E_{\eta}^{+,h}(\textbf{k}),T_{h})$. See Eq.~13 for the definition of Fermi function.}
    	\label{fig:Otto_cycle}
    \end{figure}

    The $4$-stroke cycle of a  QOE involves two quantum isochoric and two quantum adiabatic strokes as shown in Fig~\ref{fig:Otto_cycle}. In the stage $A \rightarrow B$, the system is kept in contact with hot bath at temperature $T_{h}$. In this process heat flows into the system until the working substance is in thermal equilibrium with the heat bath. As the energy eigenvalues remain unchanged during the quantum isochoric process~\cite{quan2007quantum}, photo illumination parameter is set at $I_{+}^{h}$ during $A \rightarrow B$. In stage $B \rightarrow C$, we have a quantum adiabatic process which means no heat transfer and this implies constant occupation number($P_{+}(T_{h})$)~\cite{quan2007quantum} in the quantum case. During this step, the system is kept isolated from the two thermal baths.
      The process $C \rightarrow D$ is again a quantum isochoric process with illumination parameter set at $I_{+}^{c}$ and on this occasion, the system is kept in contact with cold bath having temperature $T_{c}$. During this process, heat flows out of the system so as to reach thermal equilibrium with the cold bath. Finally, the adiabatic process $D  \rightarrow A$ at constant occupation number $P_{+}(T_{c})$ completes the cycle. The positive band contribution to input and output heat is given as
    \begin{eqnarray}
        Q_{in}^{+} &=& 2\sum_{\eta}\int \frac{d^{2}\textbf{k}}{(2\pi)^{2}} E_{\eta}^{+,h}(\textbf{k})(P_{+}(T_{h})-P_{+}(T_{c})),
        \label{eqn:qin}\\
        Q_{out}^{+} &=& 2\sum_{\eta}\int \frac{d^{2}\textbf{k}}{(2\pi)^{2}}E_{\eta}^{+,c}(\textbf{k})(P_{+}(T_{c})-P_{+}(T_{h}))
        \label{eqn:qout}
    \end{eqnarray}
    where $E_{\eta}^{+,h}$ is the energy for the isochoric process $A \rightarrow B$ with illumination parameter $I_{+}^{h}$ and $E_{\eta}^{+,c}$ is the energy of the isochoric process $C \rightarrow D$ with illumination parameter $I_{+}^{c}$. Due to re-normalization, we just need to concentrate on positive bands. The total heat input and output is given by $Q_{in} = 2Q^{+}_{in}$ and $Q_{out} = 2Q^{+}_{out}$ where $+$ represents the positive band contribution. We know that for quasisteady epitaxial graphene there is only one positive band with degeneracy two due to spin. But we don't have any valley degeneracy like stanene~\cite{fadaie2018topological} and this requires separate treatment of $K$ and $K^{\prime}$ points,i.e. we have $Q^{+} = Q^{+}_{K} + Q^{+}_{K^{\prime}}$ both for input and output heat. The $\eta$ sum over $\{K,K^{\prime}\}$ in Eq.~\ref{eqn:qin} $\&$ \ref{eqn:qout} takes care of this fact. The net work output $W_{O} = 2W_{O}^{+}$ where the positive band contribution is given by
    \begin{eqnarray}
    &&W_{O}^{+} = Q_{in}^{+} + Q_{out}^{+}\nonumber\\&& = 2\sum_{\eta}\int \frac{d^{2}\textbf{k}}{(2\pi)^{2}}(E_{\eta}^{+,h}(\textbf{k}) - E_{\eta}^{+,c}(\textbf{k}))(P_{+}(T_{h}) - P_{+}(T_{c}))
    \label{eqn:positve_otto_work}
    \end{eqnarray}
    and the efficiency of QOE cycle is $\eta_{o}( = W_{O}/Q_{in}$).
    
  \subsubsection{\rm{\textbf{Quantum Stirling engine cycle}}}
\begin{figure}[H]
    	\begin{center}
    		\includegraphics[width=7.25cm,height=6.25cm]{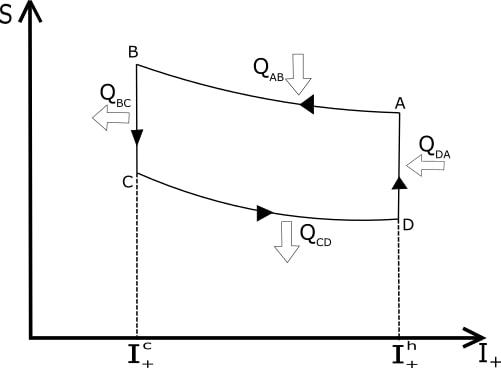}
    	\end{center}
    	\caption{Entropy(S) vs illumination parameter($I_{+}$). QSE cycle has two isothermal and two isochoric processes- $A\rightarrow B$, $C \rightarrow D$ and $B\rightarrow C$, $C \rightarrow D$ respectively. The heat energy input in the $A \rightarrow B$ and $D \rightarrow A$ stages are $Q_{AB}$ and $Q_{DA}$ respectively. $Q_{BC}$ and $Q_{CD}$ are the heat energy output in $B \rightarrow C$ and $C \rightarrow D$ stages. The hot and cold bath temperatures for the QSE cycle are set at $T_{h}$ and $T_{c}$ respectively. The illumination parameter is varied between $I_{+}^{c}$ and $I_{+}^{h}$.}
    	\label{fig:Stirling}
    \end{figure}

    The $4-$stroke cycle of the QSE~\cite{fadaie2018topological,ma2017quantum,thomas2019quantum,huang2014quantum} consists of two quantum isochoric (or isoenergetic) and two quantum isothermal processes as shown in Fig~\ref{fig:Stirling}. The stages $A \rightarrow B$ and $C \rightarrow D$ are the isothermal processes with associated heat bath temperature $T_{h}$ and $T_{c}$ respectively. The stages $B \rightarrow C$ and $D \rightarrow A$ are two isochoric processes in the cycle.  
        The positive band contribution to heat flow at different stages of the quantum Stirling engine cycle is given as $Q_{AB}^{+} = T_{h}(S^{+}(B) - S^{+}(A))$, $Q_{BC}^{+} = U^{+}(C) - U^{+}(B)$, $Q_{CD}^{+} = T_{c}(S^{+}(D) - S^{+}(C))$, and $Q_{DA}^{+} = U^{+}(A) - U^{+}(D)$ with $U^{+}(i)$ and $S^{+}(i)$ being internal energy and entropy at instants $i \in \{A, B, C, D\}$ for the positive band. $U^{+}(i)$ is calculated from Eq.~\ref{renormalized_internal_energy} and the explicit form of entropy in terms of occupation probability is given by 
        \begin{eqnarray}
            S^{+}(i) &=& -2k_{B}\int \frac{d^{2}\textbf{k}}{(2\pi)^{2}}\sum_{\eta}\{(1-P_{+}(T_{i}))\ln(1-P_{+}(T_{i}))\nonumber\\ &+& P_{+}(T_{i})\ln P_{+}(T_{i})\}.
        \label{eqn:entropy}
        \end{eqnarray} Total heat at different stages of the cycle are given by $Q_{AB} = 2Q^{+}_{AB}$, $Q_{BC} = 2Q^{+}_{BC}$, $Q_{CD} = 2Q^{+}_{CD}$, and $Q_{DA} = 2Q^{+}_{DA}$. The factor two arises because of renormalization of Internal energy. The work output in one complete cycle of QSE is
        \begin{eqnarray}
            W_{S} = Q_{AB} + Q_{BC} + Q_{CD} + Q_{DA}
            \label{eqn:stirling_work_formula}
        \end{eqnarray}
        On adding the heat at each stage of the cycle we have four terms like $T_{i}S^{+}(i) - U^{+}(i)$ where $T_{i}$ is the temperature at instant $i$, each of these terms is the negative of grand canonical potential~\cite{pathria32statistical} and this becomes free energy for distinguishable particles, thus we get
        \begin{eqnarray}
            W_{S} = 2k_{B}\int\frac{d^{2}\textbf{k}}{(2\pi)^{2}}\left[T_{h}\ln\frac{\mathcal{Z}(B)}{\mathcal{Z}(A)} + T_{c}\ln\frac{\mathcal{Z}(D)}{\mathcal{Z}(C)}\right],
            \label{eqn:work_stirling}
        \end{eqnarray}
        where $\mathcal{Z}(i)(=\Pi_{\eta}\{1+\exp (-E_{\eta}^{+,i}(\textbf{k})/k_{B}T_{i})\}^{2})$ is the partition function~\cite{pathria32statistical,fadaie2018topological} at instant $i$, $E_{\eta}^{+,i}(\textbf{k})$ is the energy eigenvalue at instant $i$, while $k_{B}$ is the Boltzmann constant. The efficiency of the QSE is given by ratio of work output and input heat $Q_{in} = Q_{AB} + Q_{DA}$, i.e. $\eta_{S} = W_{S}/Q_{in}$.  
\section{Results and Discussion}
   \subsection{Quantum Otto engine cycle}    
    For calculating the work output and efficiency, we fix the low temperature regime heat baths at $T_{c} = 30$ K and $T_{h} = 40$ K while for high temperature regime, heat baths are set at $T_{c} = 150$ K and $T_{h} = 300$ K. 
    \begin{figure}[H]
        \centering
        \subfigure{\includegraphics[width=0.35\textwidth,height = 6cm]{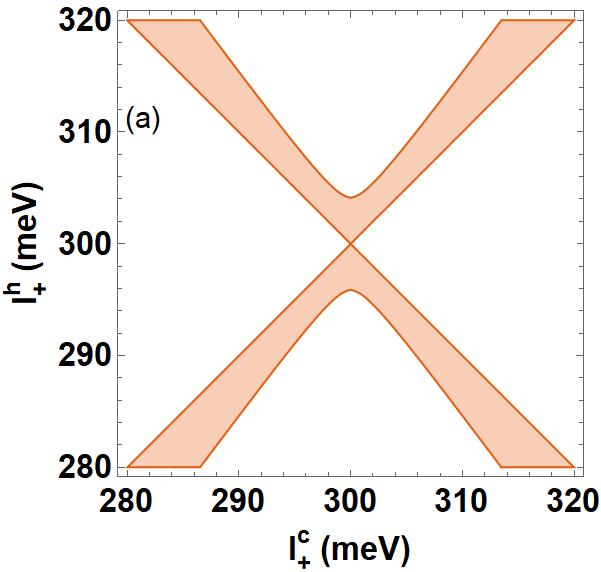}\label{fig:otto_work_region_plot_low_temp}} 
        \subfigure{\includegraphics[width=0.35\textwidth,height =6cm]{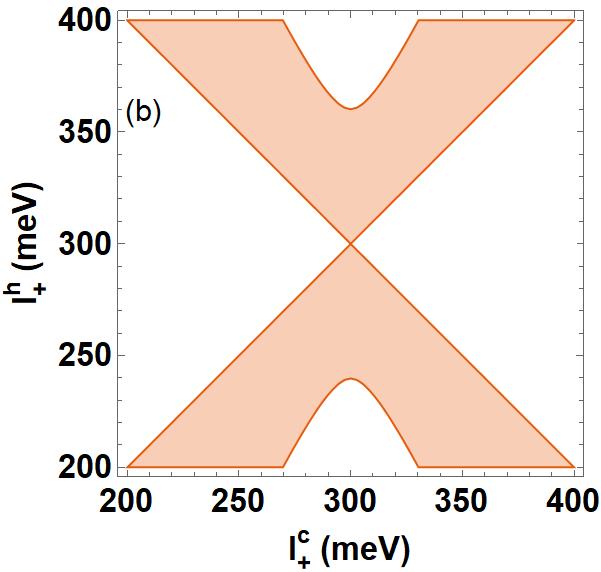}\label{fig:otto_work_region_plot_high_temp}} 
        \caption{Positive work region in the QOE cycle. The orange region shows the positive work output and the remaining portion indicate negative work output. The substrate potential $\Delta$ is set at $300$ meV and the hot and cold bath temperatures are (a) $T_{c} = 30$ K and $T_{h} = 40$ K and (b) $T_{c} = 150$ K and $T_{h} = 300$ K.}
    \end{figure}
    The positive work region is shown in Fig.~\ref{fig:otto_work_region_plot_low_temp} and \ref{fig:otto_work_region_plot_high_temp} for low and high temperature regimes.~We get nice $X$-shaped structures where orange region denotes positive work and vacant (white regions) regions of the plot are for negative work which can be interpreted as refrigerator cycle or heat pump cycle. We notice that the $X$-shaped structure is not destroyed even at high temperature which is markedly distinct for the case depicted in Ref.~\cite{fadaie2018topological} as it takes a shape which is not symmetric about the phase transition point.~The symmetry of $X$-shape in case of the epitaxial graphene based FTI about phase transition point ($\Delta= I_{+}^{c} = I_{+}^{h} = 300$ meV) suggests that BI and FTI phases are equivalent in terms of work output. In case of  FTI, the X shaped structure gets broader at high temperatures but the symmetry is still preserved about the topological phase transition point.  
    
    During the QOE cycle of Fig.~\ref{fig:Otto_cycle}, the illumination parameter is varied between $I_{+}^{c}$ and $I_{+}^{h}$. We present plots for work output versus illumination parameter ($I_{+}^{c}$) for three distinct values of illumination parameter ($I_{+}^{h}$). The efficiency as a function of $I_{+}^{c}$ for the same three values of $I_{+}^{h}$ is also shown below their corresponding work output plot of QOE. 
    \begin{figure}[H]
        \subfigure{\includegraphics[width=6.5cm,height=5cm]{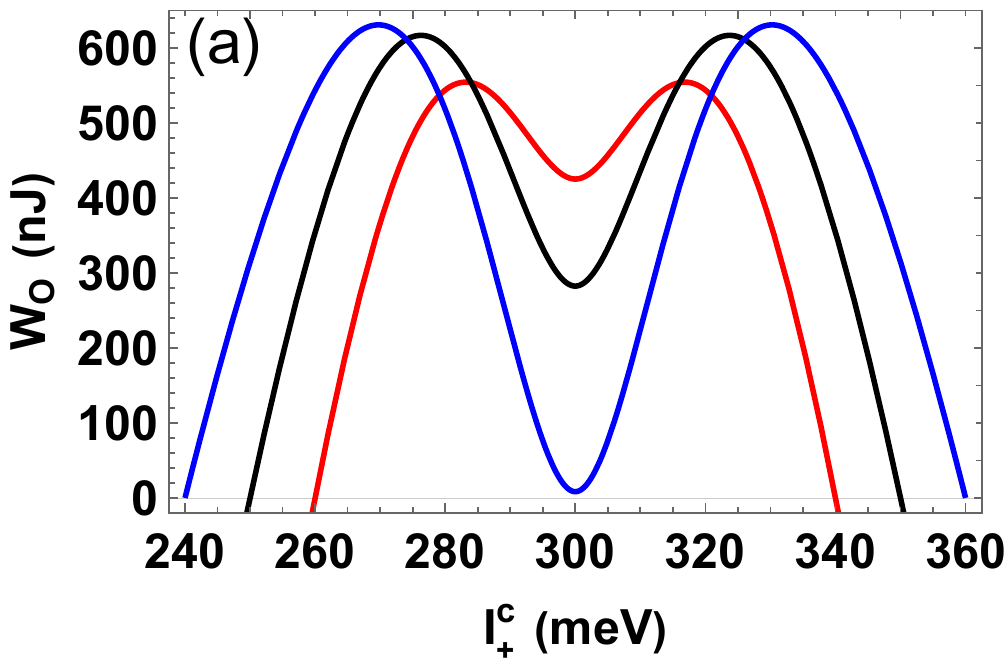} \label{fig:work_otto_high_temp}}
    	\subfigure{\includegraphics[width=8.5cm,height=5cm]{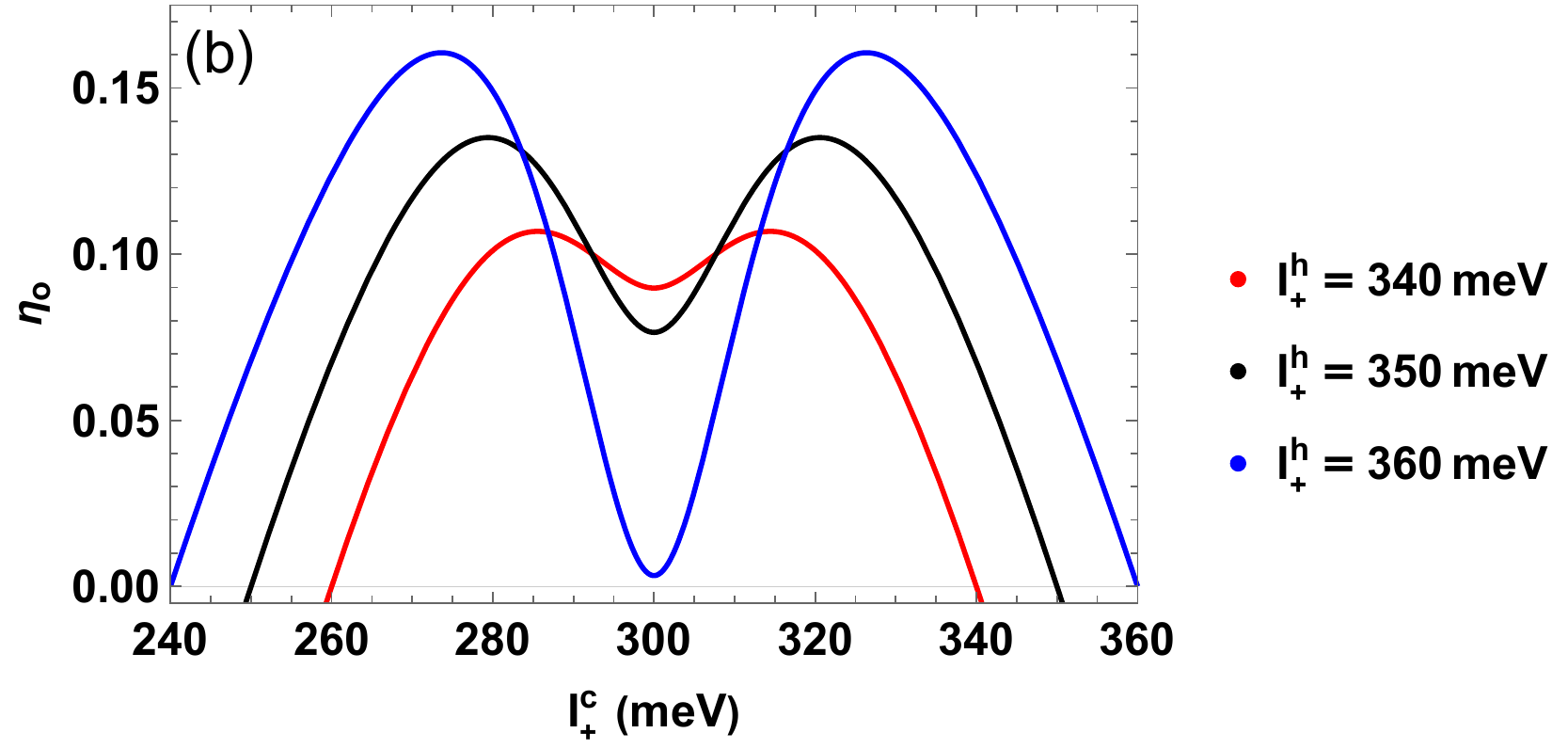} \label{fig:efficiency_otto_high_temp}}
    	\vskip -0.5cm
    	\caption{(a) Work output $W_{O}$ (in units of nano Joule) and (b) Efficiency $\eta_{o}$ of the QOE cycle are plotted versus $I_{+}^{c}$ for three different values of $I_{+}^{h} = 340$ meV(red), 350 meV(black), and 360 meV(blue). The temperatures of the cold and hot bath are $T_{c} = 150$ K and $T_{h} = 300$ K both for (a) and (b).~{The local minima at $I_{+}^{c} = 300$ meV captures the topological transition point with  $I_{+}^{c} < 300 meV$ being BI regime while $I_{+}^{c} > 300 meV$ is FTI regime.}}
    	\label{fig:otto_high_temp}
    \end{figure}
    {In Fig.~\ref{fig:work_otto_high_temp} we plot the work output and in Fig.~\ref{fig:efficiency_otto_high_temp} the efficiency of QOE cycle in the high temperature regime. In all cases for the work output and efficiency, we see a double peak structure with local minima at the phase transition point, $\Delta = 300 meV$, with $I_{+}^{c} > \Delta$ defines the FTI regime and $I^{c}_{+} < \Delta$ defining the BI regime.~Both work output and efficiency plots look similar as they are related by the relation $\eta_{O}=W_{O}/Q_{in}$ and the heat input $Q_{in}$ remain symmetric about the transition point.~Also the spacing between $Q_{in}$ values for any two given values of $I_{+}^{h}$ remain same nearly for all $I^{c}_{+}$ values plotted in Fig.~7(a) and 7(b).~The simultaneous occurrence of double peak for work output and efficiency plots signifies that we can extract maximum work from the QOE with maximum efficiency.} If illumination parameter $I_{+}^{h}$ is set at $240$ meV, $250$ meV, and $260$ meV, the work output and efficiency curve in Fig.~\ref{fig:otto_high_temp} is replicated. This suggests that phases(BI and FTI) appear to be same in terms of work output and efficiency of QOE at high temperatures. Nevertheless, the topological phase transition point can still be probed in all cases since we get an extremum at $I_{+}^{c} = 300$ meV for both work output and efficiency plots. There is a pronounced dip in both the work output and efficiency.
    \begin{figure}[H]
        \subfigure{\includegraphics[width=6.5cm,height=5cm]{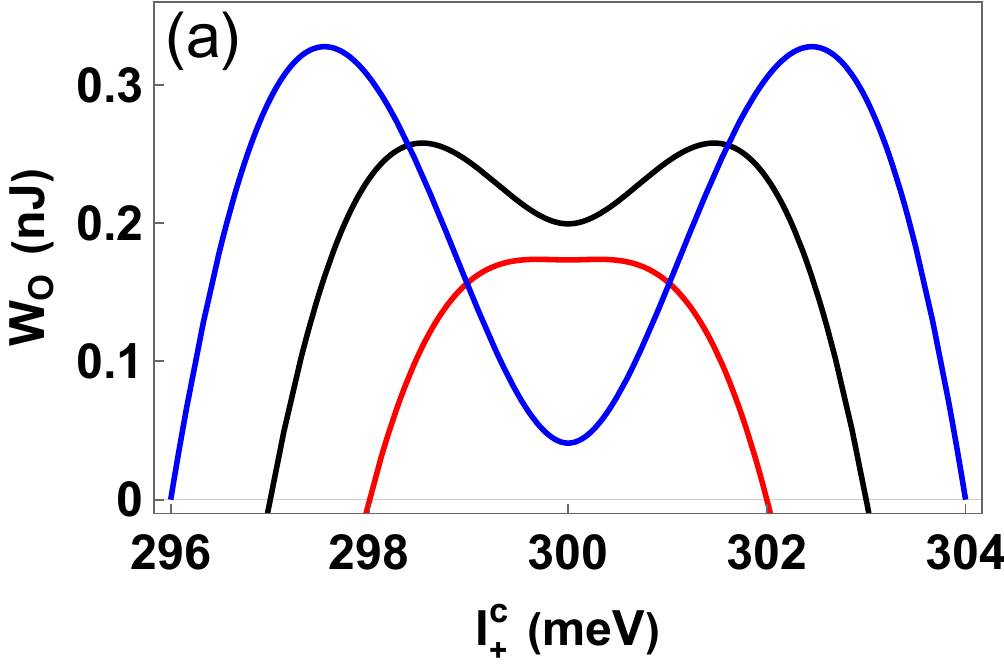} \label{fig:work_otto_low_temp}}
    	\subfigure{\includegraphics[width=8.5cm,height=5cm]{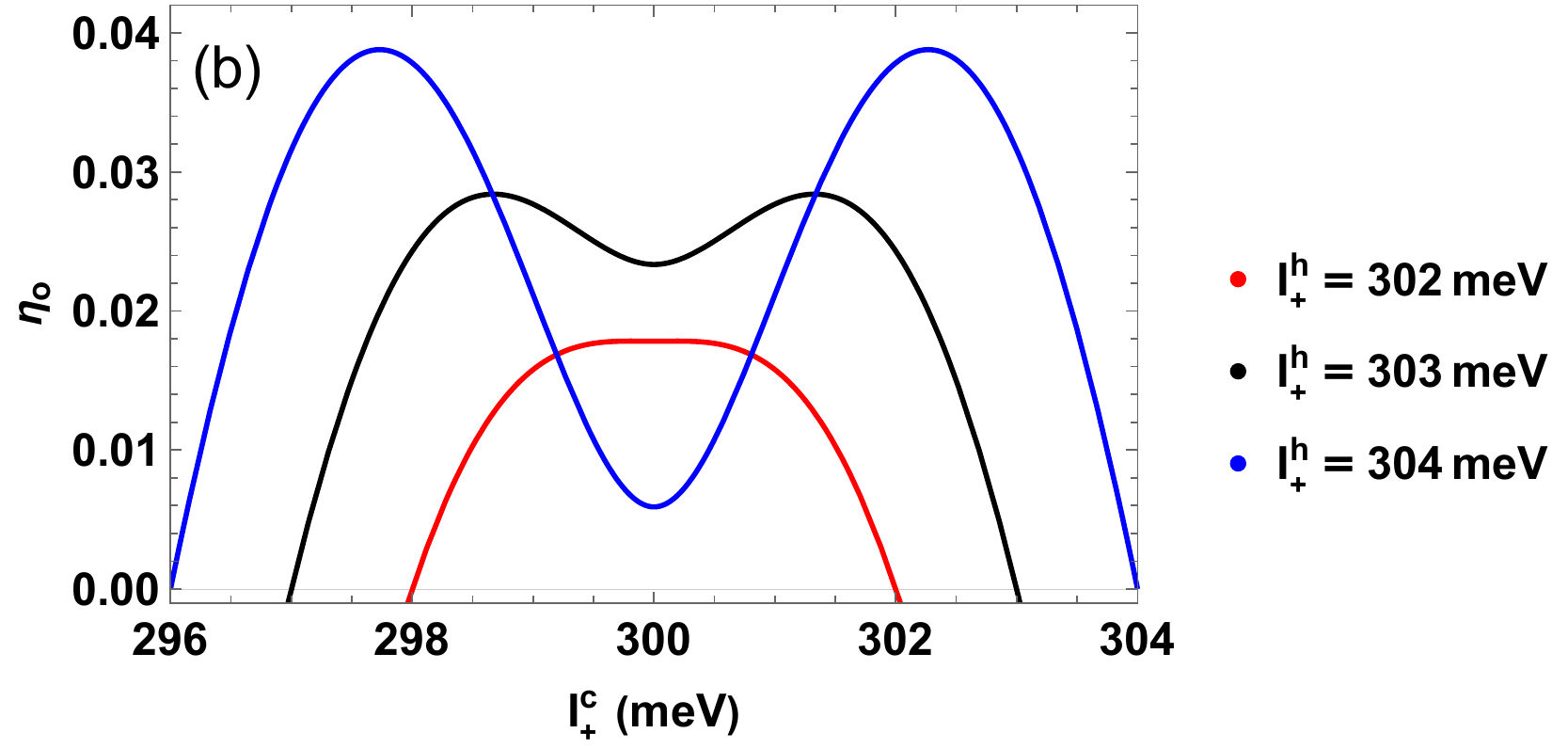} \label{fig:efficiency_otto_low_temp}}
    	\centering
    	\caption{(a)~Work output $W_{O}$(in units of nano Joule) and (b) Efficiency $\eta_{o}$ of the QOE cycle are plotted as a versus $I_{+}^{c}$ for three distinct values of $I_{+}^{h} = 302$ meV(red), $303$ meV(black), and $304$ meV(blue). The temperatures of the cold and hot bath are $T_{c} = 30$ K and $T_{h} = 40$ K both for (a) and (b).~{The local minima at $I_{+}^{c} = 300$ meV captures the topological transition point with $I_{+}^{c} < 300 meV$ being BI regime while $I_{+}^{c} > 300 meV$ is FTI regime.}}
        \label{fig:otto_low_temp1}
    \end{figure}
   {Fig.~\ref{fig:work_otto_low_temp} and \ref{fig:efficiency_otto_low_temp} shows work output and efficiency at low temperature with $T_{c}$ = 30 K and $T_{h}$ = 40 K. The work output is in orders of few nano Joules and efficiency drops substantially as compared to the high temperature QOE. An extremum can be seen at the phase transition point for the work and efficiency plots.~In all cases for the work output and efficiency, we see a double peak structure with local minima at the phase transition point, $\Delta = 300 meV$, with $I_{+}^{c} > \Delta$ defines the FTI regime and $I^{c}_{+} < \Delta$ defining the BI regime.~Both work output and efficiency plots look similar as they are related by $\eta_{O}=W_{O}/Q_{in}$ and the heat input $Q_{in}$ remain symmetric about the transition point.~Also the spacing between $Q_{in}$ values for any two given values of $I_{+}^{h}$ remain same nearly for all $I^{c}_{+}$ values considered in Fig.~8(a) and 8(b).~The simultaneous occurrence of double peak for work output and efficiency plots signifies that we can extract maximum work from the QOE with maximum efficiency.}
    
    The above plots depicted in Fig.~\ref{fig:otto_high_temp} and~\ref{fig:otto_low_temp1} remain same even if we change the direction of right circularly polarized light to left. This can be realized by looking at Eqs.~(\ref{eqn:qin}-\ref{eqn:positve_otto_work}). The change in polarization direction, leads to a switch of $K$ with $K^{\prime}$ in the spectrum shown in Fig.~\ref{fig:spectrum}. This switching does not affect the expressions for $Q_{in}^{+}$, $Q_{out}^{+}$, and $W_{O}^{+}$.~Thus measurement of both work and efficiency in a QOE is robust against the direction of polarization of circularly polarized light. 

\subsection{Quantum Stirling engine cycle}
\begin{figure}[H]
        \subfigure{\includegraphics[width=6.5cm,height=5cm]{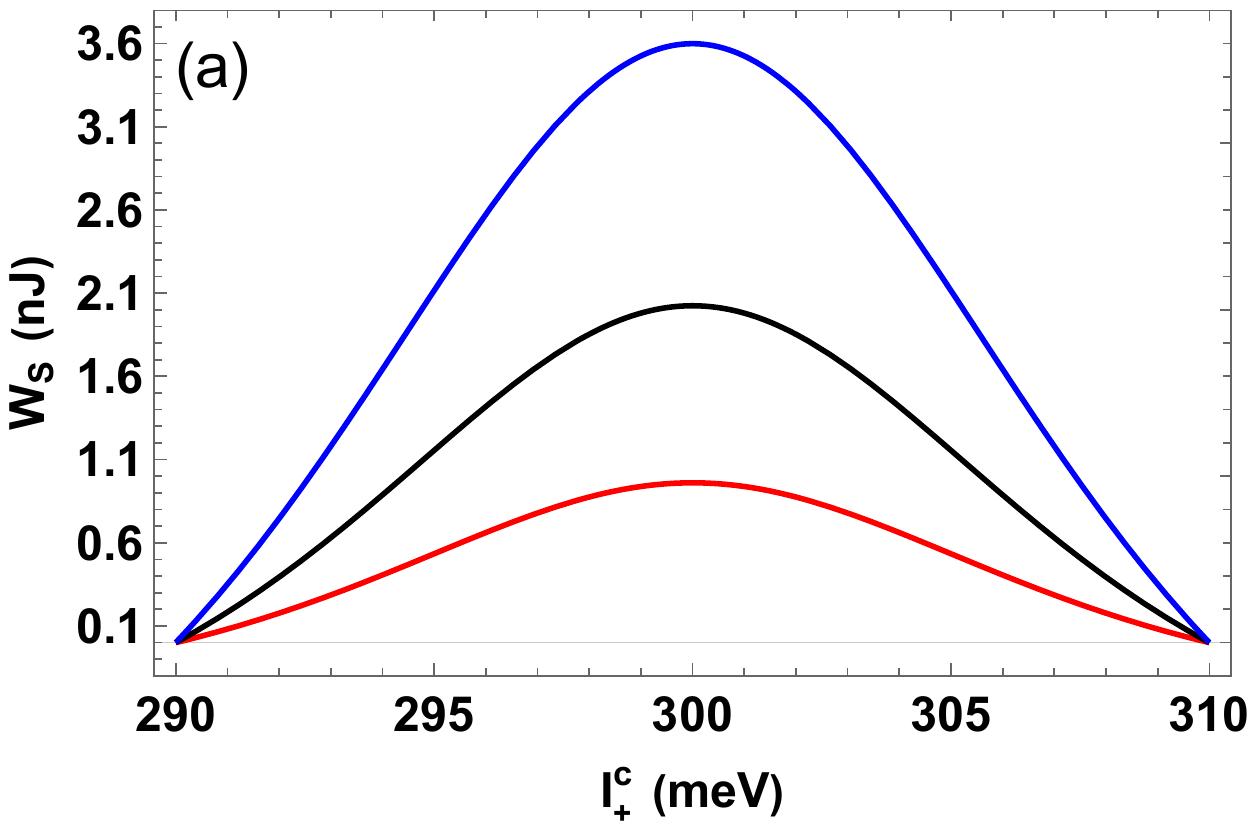} \label{fig:work_stirling_low_temp}}
    	\subfigure{\includegraphics[width=8.5cm,height=5cm]{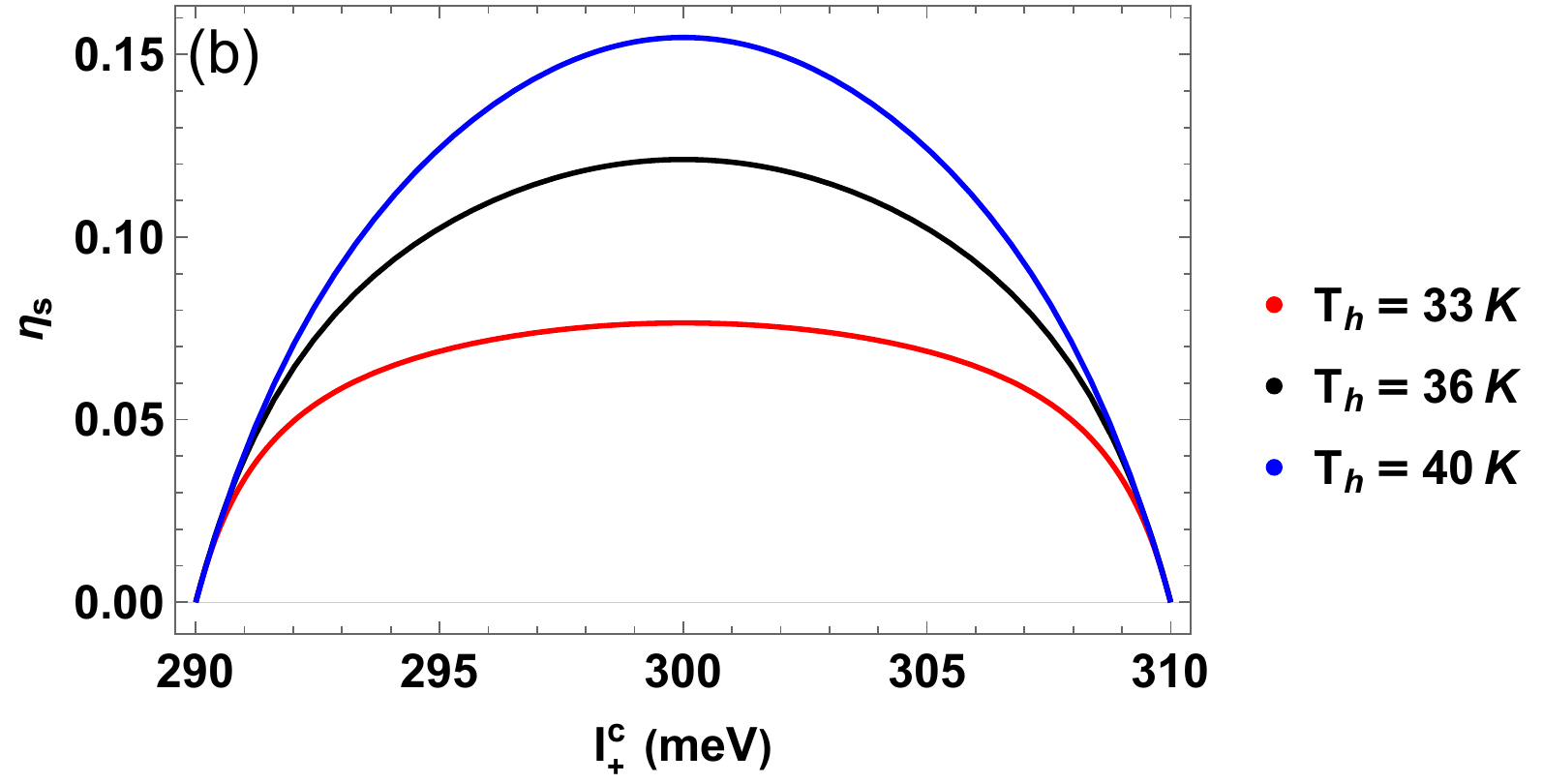} \label{fig:efficiency_stirling_low_temp}}
    	\centering
    	\caption{(a)~Work output $W_{S}$(in units of nano Joule) and (b) Efficiency $\eta_{s}$ of the QSE cycle are plotted as a function of $I_{+}^{c}$ for three different values of $T_{h} = 33$ K(red), 36 K(black), and 40 K(blue). The temperatures of the cold bath is set at $T_{c} = 30$ K and the illumination parameter $I_{+}^{h} = 310$ meV.~{The maxima at $I_{+}^{c} = 300$ meV captures the topological transition point with $I_{+}^{c} < 300 meV$ being BI regime while $I_{+}^{c} > 300 meV$ is FTI regime.}}
    	\label{fig:work_efficiency_stirling_low_temp}
    \end{figure}

      The QSE work output and efficiency for the low temperature regime is plotted in Fig.~\ref{fig:work_efficiency_stirling_low_temp}. The work output is in the range of few nano Joules and positive work output region becomes broader with increase in hot bath temperature $T_{h}$. The maximum of work output and efficiency occurs each time at phase transition point $I_{+}^{c} = 300$ meV and for $I_{+}^{h} = 310$ meV. Since for both the work and efficiency plot, the maxima occurs at phase transition point. Therefore it can be easily distinguished.  
    The high temperature study of QSE for a FTI as shown in Fig.~\ref{fig:work_stirling_high_temp} is quite different from what was seen in the topological insulator QSE of Ref.~\cite{fadaie2018topological}. In Ref.~\cite{fadaie2018topological}, the topological phase transition point is smoothed out since maximum of work output is shifted away from the phase transition point on increase of hot reservoir temperature. But in Fig.~\ref{fig:work_stirling_high_temp}, one can clearly identify that maximum occurs at the phase transition point for all three temperatures of the hot reservoir. The maximum work output is in orders of $100$ nJ and it increases as we increase hot bath temperature.
    \begin{figure}[H]
        \subfigure{\includegraphics[width=6.5cm,height=5cm]{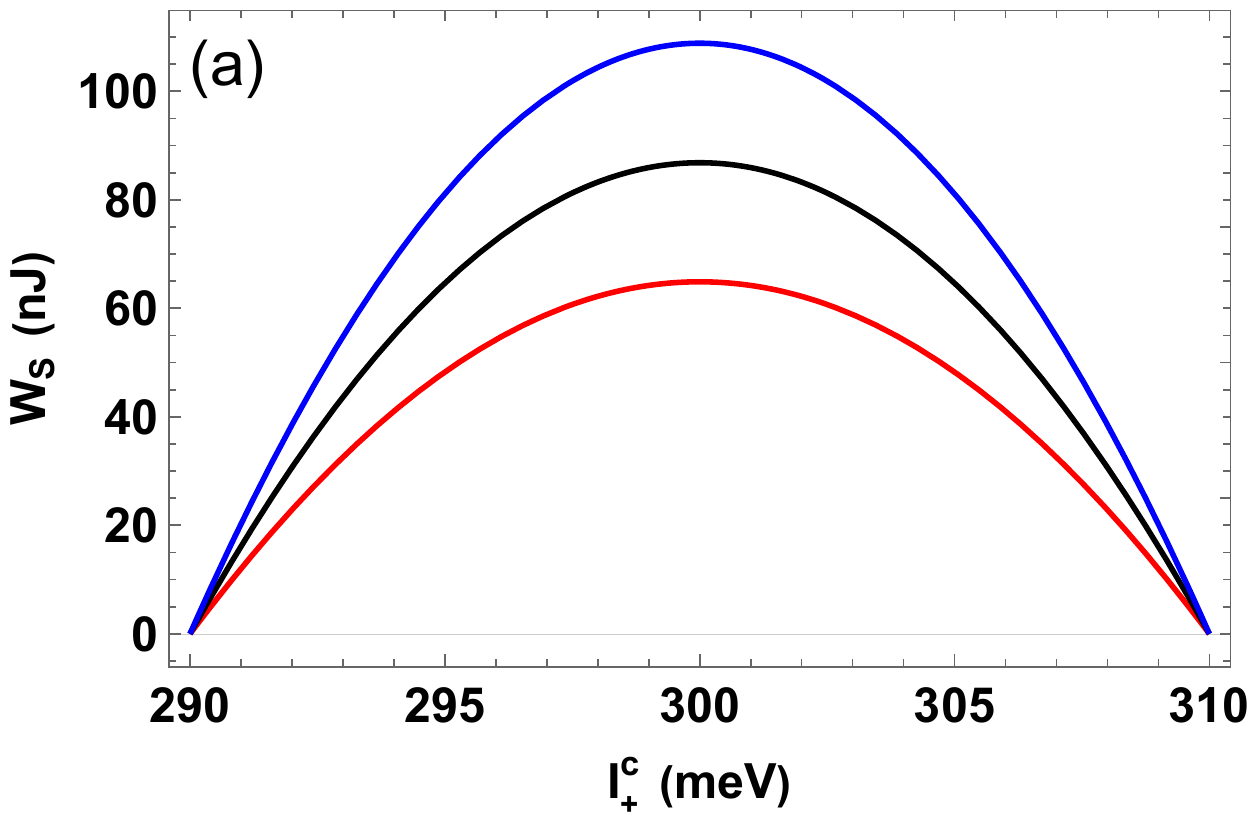} \label{fig:work_stirling_high_temp}}
    	\subfigure{\includegraphics[width=8.5cm,height=5cm]{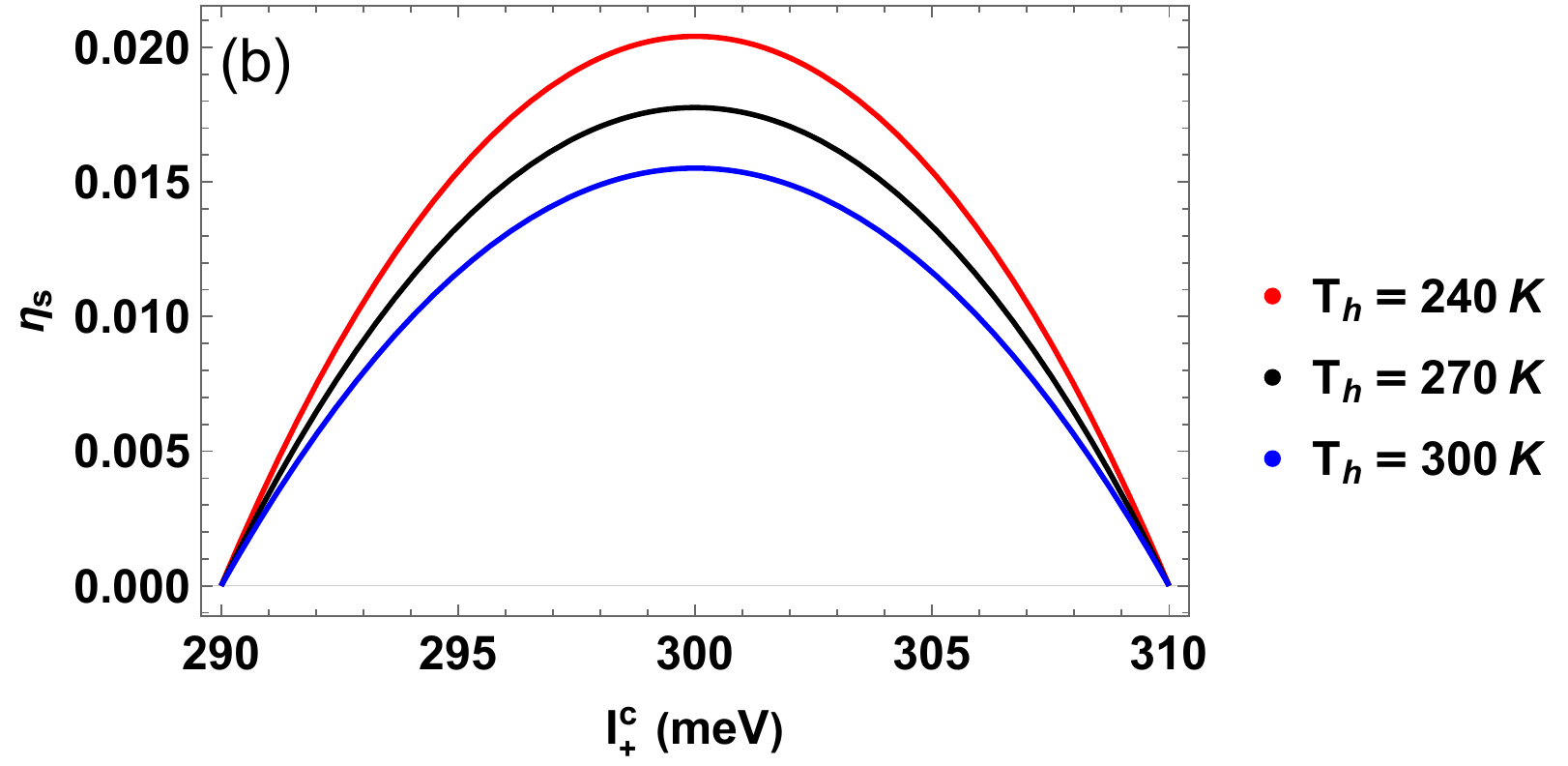} \label{fig:efficiency_stirling_high_temp}}
    	\centering
    	\caption{(a)~Work output $W_{S}$(in units of nano Joule) and (b)~Efficiency $\eta_{s}$ of the QSE cycle are plotted as a function of $I_{+}^{c}$. The temperature of cold bath is set at $T_{c} = {150}$ K and the hot bath temperature is kept at three different values $T_{h} = 240$ K(red), $270$ K(black), and $300$ K(blue). The illumination parameter $I_{+}^{h}$ is set at $310$ meV.~{The maxima at $I_{+}^{c} = 300$ meV captures the topological transition point with $I_{+}^{c} < 300 meV$ being BI regime while $I_{+}^{c} > 300 meV$ is FTI regime.}}
    	\label{fig:work_stirling_high_temp}
    \end{figure}    
     The efficiency is plotted for both low and high temperature regimes in Figs.~\ref{fig:efficiency_stirling_low_temp} and~\ref{fig:efficiency_stirling_high_temp} where we set $I_{+}^{h} = 310$ meV. Both low and high temperature regime plots reach their maximum at the phase transition point.~The symmetry about topological phase transition point is present in all the curves discussed for QSE. We dealt with right circularly polarized light in the computation shown above but we get exactly same result on using left circularly polarized light. As changing polarization direction is equivalent to switching $K$ and $K^{\prime}$ point in the spectrum, and work output and heat input expressions for QSE remain invariant to this change.~Hence this makes work output and efficiency of a FTI-QSE cycle robust against the direction of polarization of circularly polarized light. 
\section{Experimental realization}
    A substrate induced tunable band gap has been realized in experiments on monolayer expitaxial graphene~\cite{zhou2007substrate,quhe2012tunable}. Recently, experiments with continuous time dependent and pulsed laser were done to demonstrate new quantum phases present in graphene~\cite{zhang2020light,mciver2020light}.~To construct a thermodynamic cycle for quantum Otto and Stirling engines with epitaxial graphene as a working substance, cold and hot thermal baths should be periodically kept in contact with FTI to drive the quantum isochoric and isothermal processes in the heat engine.~{For both the engines, epitaxial graphene is kept in contact with hot reservoir for the thermodynamic process between $A\rightarrow B$ and with cold reservoir for the process between $C\rightarrow D$.~A three dimensional graphene architecture, see Fig.~6 of Ref.~\cite{pop2012thermal}, or bilayer/ multilayer graphene~\cite{zhang2016robustly,rajabpour2012tuning} can act as an optimal tunable heat bath for the light induced epitaxial graphene.}~Laser emitting a circularly polarized sinusoidal light should be used to control the photo-illumination parameter during the cycle.~The ultrahigh thermal conductivity of graphene makes it an ideal choice for experiments with  high frequency light~\cite{pop2012thermal}.~The topological properties using high frequency laser on graphene have been explored both theoretically~\cite{PhysRevB.99.214302,sato2019light} and experimentally~\cite{sentef2015theory,mciver2020light} at temperatures near 100~$K$.~These studies take care of the relaxation time corresponding to electron thermalization and electron scattering time.~The heating of sample due to laser will be suppressed remarkably by working in the off-resonant frequency ($\omega \gg 10^{15} Hz$) regime~\cite{rudner2020band}.~We fix the substrate potential $\Delta$ to be $300$~$meV$ where we see the numerical work output and efficiency results clearly but it can be fine tuned to any value in the range $150-350$~$meV$ using a gate voltage~\cite{quhe2012tunable,zhou2007substrate}.~The probe for the phase transition point in QOE and QSE cycle works well for all values of the substrate potential. {In the study of low and high temperature heat engine, our results for QOE are restricted to bath temperature difference ($T_{h}-T_{c}$) of $10$~$K$ and $150$~$K$.~We would like to mention here that the stated results hold true even for temperature difference near $1$~$K$ for both low and high temperature regimes of the Otto cycle.~However the reason to choose such bath temperatures in the article is to show noticeable enhancement of work output~(by a factor of 100) with increase in bath temperature difference from $10$~$K$ to $150$~$K$.~We notice that below a temperature difference of 1~$K$ the work output become negative.~Since our article is a study using heat engine, therefore we have not incorporated the results of a refrigerator with negative work output.~Further the work output for a refrigerator also have extremum at topological phase transition point.~The above statements also hold true for the proposed cycle of QSE.~Thus regardless of the bath temperature difference being high or low, we can always obtain an extremum near topological phase transition.}~This flexibility in experimentally allowed values for substrate potential, bath temperatures, and the corresponding illumination parameter makes probing the phase transition point via the design of such quantum heat engines, perhaps an ideal choice to study rich Floquet topological phases. {The experiment with pump-probe time resolved, angle resolved photoemission spectroscopy~(tr-ARPES) extract band structure details (band crossing or gap closing) to realize Floquet topological phases in graphene~\cite{sentef2015theory}.~However they use pulsed laser to mimic the effect of circularly polarized light.~Our method provide an alternative route to probe these phases using a thermodynamic heat engine cycle with continuous time dependent circularly polarized light. It not only provides information of different topological phases but can also extract large work output in units of nanoJoule. Together with a probe it also act as a nano scale heat engine which can help to extract work output at reasonable efficiency. Thus our method of detecting topological phases via a quantum thermodynamic cycles scores over a tr-ARPES method which only acts as a probe while our method works both as a probe as well as a quantum device generating electric power via the thermodynamic cycle.} 

\section{Conclusion}
   We have detected the photoinduced phase transition point in FTI using work and efficiency of QOE and QSE cycles. The work output and efficiency takes a two peak structure for QOE and there is an extremum at the phase transition point. For QSE, we notice single peak structure in work and efficiency plots where maxima occurs at the phase transition point.~The use of work output and efficiency as a probe of the phase transition point is not limited to low temperature only but can also be seen for high temperatures for both QOE and QSE cycle.~Earlier in Ref.~\cite{fadaie2018topological} it had been pointed out that QHE cycles used to probe the topological phase transition point were only good at low temperatures.~However as we show in this manuscript QHE cycle used to probe topological phase transition point can work both at low as well as high temperature regimes.~Both QOE and QSE cycles are robust against the change in direction of circularly polarized light from left to right.~With the recent experimental advances in photonic Floquet phases~\cite{oka2019floquet,raposo2019evidence}, our model of photoinduced FTI based quantum heat engine makes a good choice to probe the topological phase transition point.
   \acknowledgments 
   This work was supported by the grants: 1. Josephson junctions with strained Dirac materials and their application in quantum information processing, SERB Grant No. CRG/20l9/006258, and 2. Nash equilibrium versus Pareto optimality in N-Player games, SERB MATRICS Grant No. MTR/2018/000070.
   \section{Data Availability statement}The data that supports the findings of this study are available within the article

\bibliographystyle{ieeetr}
\bibliography{reffile}
\end{document}